# Observing imperfection in atomic interfaces for van der Waals heterostructures

*Aidan. P. Rooney[1], Aleksey Kozikov[2,3,4], Alexander N. Rudenko[5] , Eric Prestat[1], Matthew J Hamer[2,3,4], Freddie Withers[6], Yang Cao[2,3,4], Kostya S. Novoselov[2,3,4], Mikhail I. Katsnelson[5], Roman Gorbachev[2,3,4]\* Sarah J. Haigh[1,4]\**

1. School of Materials, University of Manchester, Manchester M13 9PL, UK

2. Manchester Centre for Mesoscience and Nanotechnology, University of Manchester, Manchester M13 9PL, UK

3. School of Physics and Astronomy, University of Manchester, Oxford Road, Manchester, M13 9PL, UK

4. National Graphene Institute, University of Manchester, Manchester M13 9PL, UK

5. Institute for Molecules and Materials, Radboud University, 6525 AJ Nijmegen, Netherlands

6. College of Engineering, Mathematics and Physical Sciences, University of Exeter, Exeter, Devon, EX4 4SB, UK

Abstract:

Vertically stacked van der Waals heterostructures are a lucrative platform for exploring the rich electronic and optoelectronic phenomena in two-dimensional materials. Their performance will be strongly affected by impurities and defects at the interfaces. Here we present the first systematic study of interfaces in van der Waals heterostructure using cross sectional scanning transmission electron microscope (STEM) imaging. By measuring interlayer separations and comparing these to density functional theory (DFT) calculations we find that pristine interfaces exist between hBN and $MoS_2$ or $WS_2$ for stacks prepared by mechanical exfoliation in air. However, for two technologically important transition metal dichalcogenide (TMDC) systems, $MoSe_2$ and $WSe_2$, our measurement of interlayer separations provide the first evidence for impurity species being trapped at buried interfaces with hBN: interfaces which are flat at the nanometer length scale. While decreasing the thickness of encapsulated $WSe_2$ from bulk to monolayer we see a systematic increase in the interlayer separation. We attribute these differences to the thinnest TMDC flakes being flexible and hence able to deform mechanically around a sparse population of protruding interfacial impurities. We show that the air sensitive two dimensional (2D) crystal $NbSe_2$ can be fabricated into heterostructures with pristine interfaces by processing in an inert-gas environment. Finally we find that adopting glove-box transfer significantly improves the quality of interfaces for $WSe_2$ compared to processing in air.

Manuscript:

The field of two dimensional (2D) crystals is expanding rapidly and now includes materials with a wide range of electronic properties, offering opportunities to engineer a particular bandstructure at the atomic scale by layering together exfoliated crystals[1,2]. This concept of 'van der Waals heterostructures' has facilitated the fabrication of new optical and electronic components[3,4] including novel transistors[5–8] and photovoltaic devices[9,10]. The physical and electronic properties of these novel 2D multilayer architectures will depend critically on the structure, including interfacial impurity defects. Thus the nature and purity of interfaces is key for predicting the bandstructure[11,12] and adhesion strength[13] for such systems, and hence fundamental to understanding electronic transport data, electroluminescence characteristics, interlayer diffusion and mechanical failure mechanisms[10]. Density functional theory (DFT) is a powerful technique for predicting the ground state properties of encapsulated monolayers, bilayers or bulk van der Waals crystals. Yet there is limited experimental data regarding the quality of interfaces formed between dissimilar van der Waals materials[14] and many theoretical predictions therefore assume these interfaces to be free from impurities[15,16]. The assumption of perfect interfaces is justified by the observation of mobile surface contaminants segregating into bubbles leaving large flat areas of heterostructure, which have been shown to have perfect interfaces for graphene-hBN[17]. However, experimental benchmarking data regarding the interlayer separation and purity of interfaces between dissimilar 2D crystals is hard to obtain. X-ray diffraction[18] and thermal desorption spectroscopy[19] can provide interlayer separations and binding energies for bulk systems (e.g. graphite) but are not applicable for the majority of van der Waals heterostructures. Scanning probe techniques can be used to study surface morphology, but struggle to provide information for buried interfaces[20]. Cross sectional scanning transmission electron microscope (STEM) imaging is therefore the only technique capable of providing atomic resolution interfacial data for small, heterogeneous samples[10,17,21,22]. This technique has proved invaluable in the development of traditional silicon based semiconductor electronics and is able to provide high resolution structural and chemical characterization for deep subsurface regions of a 2D heterostructure device[10,17,21,22]. Here we have applied cross sectional STEM imaging to study the nature of buried interfaces and how structural properties like flake thickness influence interfaces in van der Waals' heterostructures containing transition metal dichalcogenides (TMDC). We provide the first experimental measurement of interface quality for various thicknesses of TMDC crystal with bulk hBN. Comparison of measured interlayer separations with DFT models applying the optB88 functional[23] show good agreement with experimental data for hBN-$MoS_2$, hBN-$WS_2$ and hBN-$NbSe_2$ interfaces. However, unexpectedly large interlayer separations are found for hBN-$MoSe_2$ and hBN-$WSe_2$ devices fabricated in air, suggesting the presence of trapped defects at these interfaces. Better agreement (within 0.5 Å of DFT calculations for a pristine interface) is found when the same measurements are carried out on a hBN-$WSe_2$ interface fabricated in an inert argon glovebox environment.

All the multilayer devices characterized in this work were fabricated using mechanical exfoliation and dry transfer procedures[24,25]. In summary, this involves exfoliating 2D crystal flakes from bulk, identifying flakes of suitable thickness and layering these sequentially to create a desired architecture. Figure 1 shows such a structure in plan-view using optical microscopy and scanning electron microscopy (SEM). High angle annular dark field (HAADF) STEM intensity profiles acquired perpendicular to the interface (Figure 1c) can be analyzed to determine the interlayer separation between dissimilar crystals and hence to infer interface quality. Many functional device structures incorporate a monolayer or bilayer of TMDC encapsulated inside hBN stacks[5,10,22,24–26] and here we examine individual devices incorporating monolayer $MoS_2$; monolayer $WS_2$; monolayer $MoSe_2$; bilayer $NbSe_2$, 1 – 5 layer and bulk $WSe_2$. Fabrication was performed in air except for the case of $NbSe_2$, which is known to be air sensitive so encapsulation was performed in a pure argon glovebox environment to prevent oxygen induced degradation[22]. Glovebox fabrication was also attempted for bulk $WSe_2$ to test the importance of atmosphere during fabrication. Focused ion beam milling was used in order to extract electron transparent cross sectional lamellae from the active area of these multilayer structures[27]. HAADF STEM images of encapsulated monolayers for $MoSe_2$, $MoS_2$, $WSe_2$ and $WS_2$, as well as bilayer $NbSe_2$, are shown in Figures 1 and 2. The atomic number contrast ('Z-contrast') of the HAADF imaging mode[28] means that the TMDC planes are easily identified by their higher HAADF intensity compared to the encapsulating hBN (higher intensity is shown blue/black in the temperature scale). As in previous studies of hBN-graphene heterostructures[17] the crystal planes are observed to be atomically flat over large areas. The locations of the individual neighboring hBN planes are resolvable for the $MoSe_2$, $WSe_2$ and $WS_2$, and $NbSe_2$ allowing the hBN-TMDC interlayer separation to be directly measured. The exception to this is monolayer $MoS_2$ where the contrast of the hBN planes closest to the TMDC is not clearly resolved. DFT calculations predict that for all the structures considered here the hBN-hBN c-axis lattice spacing is identical to the bulk crystal (3.33 ± 0.08 Å) even for the atomic plane closest to the TMDC (see supplementary information section 1) in good agreement with what we observe experimentally. Hence, by extrapolation from the neighboring bulk crystal we can determine the position of the unresolved hBN basal plane closest to the encapsulated $MoS_2$ layer. To obtain the mean experimental interfacial separation data for all encapsulated crystals reported in Figure 3, spacings were compared in equivalent interfaces for separately manufactured heterostructures and found to agree within experimental error. Furthermore, no differences were observed when considering interlayer spacings both above and below the TMDC layer (see supplementary information section 2).

Figure 3 compares the mean hBN-TMDC interlayer separations measured experimentally from the STEM cross sectional images to the values predicted by DFT calculations. DFT predicts little variation in the separations for pristine interfaces across all systems studied and the experimental data is in good agreement with this for $WS_2$, $MoS_2$ and $NbSe_2$. However, $MoSe_2$ and $WSe_2$ show larger interlayer separations than DFT predicts (with the deviation being 1.5 Å for $MoSe_2$ and 0.9 Å for $WSe_2$). These differences cannot be accounted for by error in the experimental values which are of the order ± 0.5 Å. A monolayer of trapped contamination can also be ruled out as this has a minimum thickness of ~1 nm[17]. Our DFT calculations have further demonstrated that variations in separation arising from different azimuthal twist angles between flakes are less than 0.02 Å in agreement with other work (see supplementary information section 1)[29].

We conclude that our pristine model does not describe $WSe_2$- or $MoSe_2$-hBN interfaces correctly. The most feasible explanation for this marked discrepancy is that these materials have defects which are chemically fixed to the TMDC surface, are thus immune to the self-cleaning phenomenon and sterically perturb the van der Waals interface. Chemically adsorbed impurity species, such as oxygen-based functional groups, are known to protrude from the surface of TMDC flakes[30]. Lattice vacancies will likely act as preferential sites for chemisorption of a range of impurities. $WSe_2$ and $MoSe_2$ have the lowest work functions of the semiconducting TMDCs (see supplementary information section 1), which favors reactions with acceptor species and suggests that defects may form readily in such systems. Nevertheless, work function is too simplistic an argument to explain the presence of impurities in these systems; NbSe$_2$ has a higher work function than WSe$_2$ and MoSe$_2$ yet is known to degrade readily under ambient conditions and requires fabrication to be performed in an argon glove box.

To better understand the interfacial behavior, different thicknesses of encapsulated WSe$_2$ from single layer to bulk at monolayer intervals were analyzed by STEM cross sectional imaging (Figure 4a-g). Figure 4h shows that the mean hBN-WSe$_2$ interlayer separation decreases for thinner TMDC layers (for values see supplementary information Table S3). Monolayer WSe$_2$ has the smallest hBN-WSe$_2$ separation but this is still 0.9 Å larger than that expected by DFT for the pristine interface. Furthermore, theory predicts no change in lattice separation for differing TMDC thicknesses. At first consideration the observed trend is surprising. One could expect that the increased interface spacing would be largest for the monolayer material which is most susceptible to surface damage, yet here we observe the opposite behavior. However, these results are not incompatible with the presence of protruding surface defects if one considers the mechanical properties of the flakes. Thinner flakes are known to be more flexible and we hypothesize that the decreasing separation moving from bulk WSe$_2$ to monolayer is associated with the increased bending modulus of thicker flakes, which scales roughly with the square of the flake thickness[31]. Consequently, a monolayer flake will be able to deform around the adsorbed species to maximize the interfacial contact with hBN as demonstrated in Figure 4i. The population of adsorbed surface defects is estimated to be <1% as our STEM chemical analysis does not reveal any notable concentration of oxygen, nitrogen or carbon at the interface or associated with the TMDCs layers (see supplementary information section 3). However chemical species residing at buried interfaces in graphene based heterostructures have been detected with larger specimen sampling using secondary ion mass spectrometry[32] and we anticipate a range of impurity species (such as Si, O, C, N, H or a combination of these) could reside at the interface.

Our results in Figure 4 can thus be explained using the assumption that steric surface impurity defects are sparsely distributed such that they do not coincide vertically. In this situation, the monolayer crystal can flexibly deform so that an average measurement of interlayer separation will find the increased thickness introduced by the adsorbed defect shared between the two interfaces; both above and below the monolayer. Conversely, bulk flakes are stiff and will sit proud on the protruding defects such that all interfaces exhibit the full increased thickness provided by the defect (see Figure 4j). It follows that the difference between the measured interlayer separation for the bulk TMDC and the pristine DFT interface will be double the difference between the monolayer and the pristine DFT. Indeed this is precisely what we observe experimentally – with the monolayer having an increased interlayer separation of 0.9 Å while the bulk has twice the increase (1.8 Å) relative to the pristine case. To better understand the extent to which an impurity may perturb the interface for a van der Waals heterostructures we have performed DFT calculations for a defective interface (supplementary information, figure S3). To assess whether device fabrication conditions can be modified to remove interfacial defects, STEM measurements were carried out for bulk $WSe_2$ on hBN, but fabricated in an inert glovebox environment. This yielded an interlayer separation of 5.51 Å, a decrease of 1.3 Å compared to air fabrication and within experimental error of the values predicted by DFT for a pristine interface.

In conclusion, we have used cross sectional STEM analysis to experimentally measure interlayer separations between hBN and the most technologically important and widely studied TMDC materials ($WS_2$, $MoS_2$, $WSe_2$, $MoSe_2$ and $NbSe_2$). We find that hBN encapsulated $MoS_2$,

WS$_2$ and NbSe$_2$ heterostructures form pristine interfaces, while MoSe$_2$ and WSe$_2$ form interfaces with larger interlayer separations than those predicted for a perfect interface. We demonstrate that this larger than expected separation is less for monolayer and bilayer TMDCs than for flakes with three or more layers. We calculate that impurity species protruding out of the plane from a WSe$_2$ flake produce an increased separation similar to what we measure experimentally. We therefore hypothesize that the behavior we observe for MoSe$_2$ and WSe$_2$ is due to the presence of a sparse population of chemisorbed steric impurity species associated with these TMDCs. The defects increase the distance from the neighboring hBN layer by 1.5 Å and 0.9 Å for MoSe$_2$ and WSe$_2$ monolayers respectively. When the TMDC flake is very thin (monolayer or bilayer) the greater flexibility allows it to conform around protruding surface sites reducing the interlayer spacing measured experimentally. Finally, we have demonstrated improved interfacial contact for WSe$_2$ through fabrication in an inert gas environment, exhibiting interfaces which approach pristine interface spacings. Given the importance of defect species for determining free carrier behavior within van der Waals heterostructures[33], this work provides vital new experimental evidence to enable us to understand and model non-optimal device behavior, as well as to devise new fabrication strategies to remove interfacial impurities and optimize device performance.

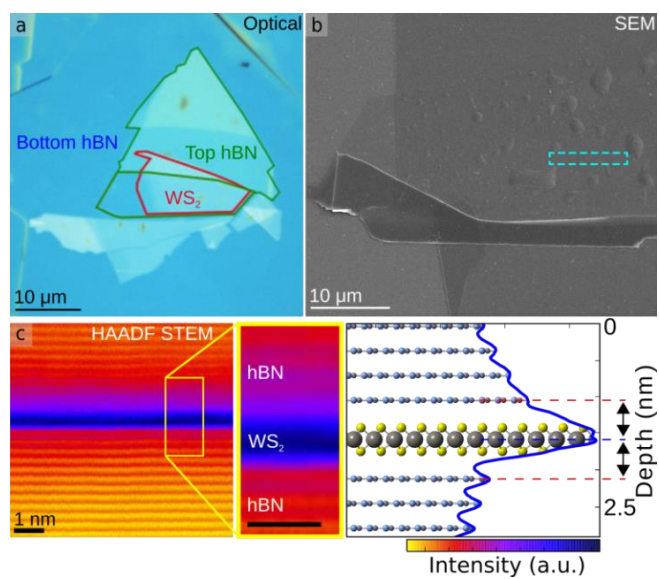

**Figure 1.** Optical, SEM and HAADF STEM imaging of $WS_2$ encapsulated in hBN. **a**, Optical plan view image of a typical heterostructure on silicon wafer. The monolayer $WS_2$ flake is highlighted red and is sandwiched between top and bottom hBN flakes, highlighted blue and green respectively. **b**, SEM plan view image of a heterostructure. Highlighted blue is a suitable region for cross sectional analysis, identified between the contamination bubbles which form as a result of self-cleaning. **c**, **Left:** HAADF STEM image of the heterostructure cross section. The image has been false colored with a non-linear scaling so that there is useful contrast in both the hBN and $WS_2$ layers. The regions between self-cleaned bubbles are atomically flat and free from mobile contaminants. **Centre:** The interfaces between $WS_2$ and hBN shown in high resolution. **Right:** A HAADF intensity profile from this image with the same scale as Centre. The highest intensity peak corresponds to the dark blue region of the image where the high atomic number $WS_2$ monolayer interacts strongly with the electron beam. The rest of the peaks in the yellow-red regions of the image are hBN monolayers. From the intensity profile the positions of the peaks can be extracted and the distance between dissimilar materials measured. The peak-peak distance shown in the annotation provides the measured interlayer separation.

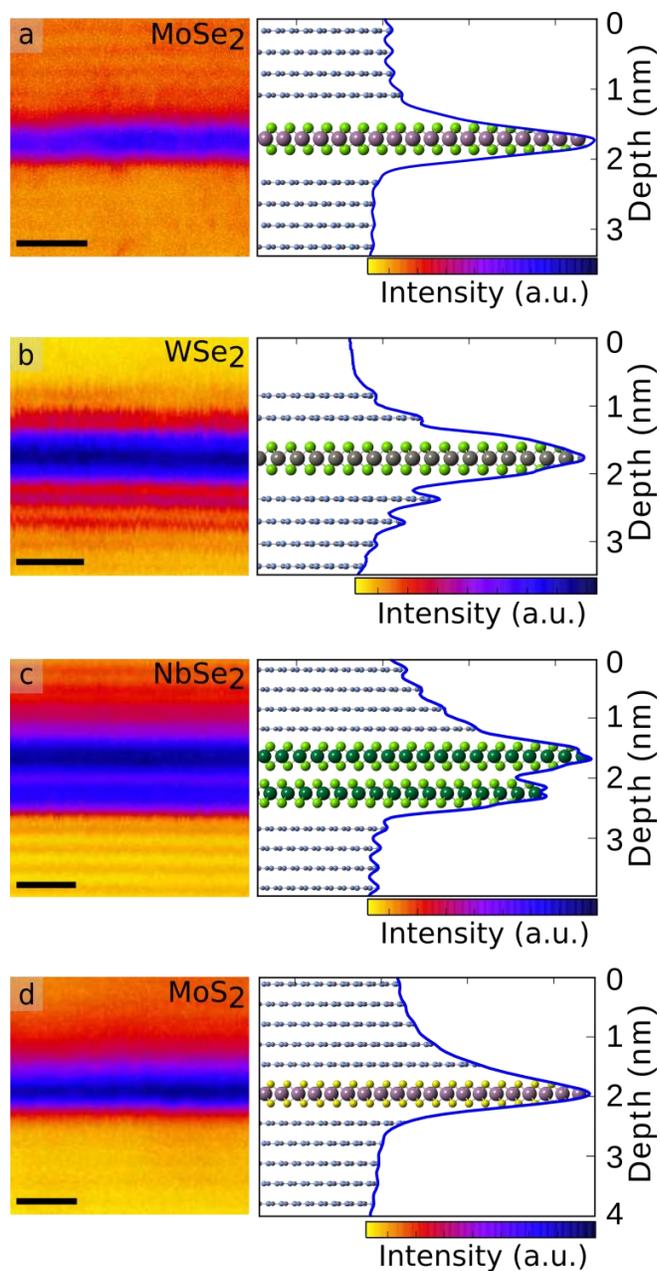

**Figure 2.** Cross sectional HAADF STEM images of different TMDCs encapsulated in hBN. **a,** monolayer MoSe$_2$; **b**, monolayer WSe$_2$; **c**, bilayer NbSe$_2$; and **d**, monolayer MoS$_2$. All scale bars 1 nm. Each image has a corresponding example intensity profile with the y-axis corresponding to the position in the HAADF image axis perpendicular to the atomic planes, and the x-axis corresponding to the HAADF intensity (indicated below as a temperature colour scale). Underlayed is an atomic model with a scale matching the intensity profile and HAADF image. The dominant peaks in the intensity profiles correspond to the heavy metals in the TMDC, and the weaker peaks to the positions of the hBN planes. Supplementary information section 2 contains further information on process of quantitatively determining interlayer separations from these images.

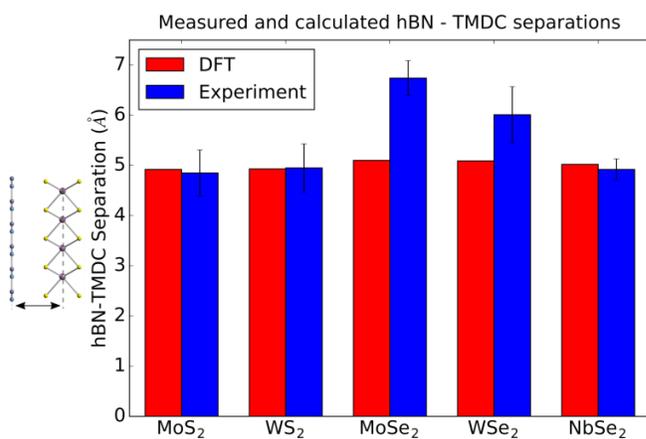

**Figure 3**. Histogram showing the mean interlayer separation values as predicted by DFT and measured by HAADF STEM. DFT and experimental values agree in almost all cases, except for $MoSe_2$ and $WSe_2$ which differ by 1.5 Å and 0.9 Å respectively.

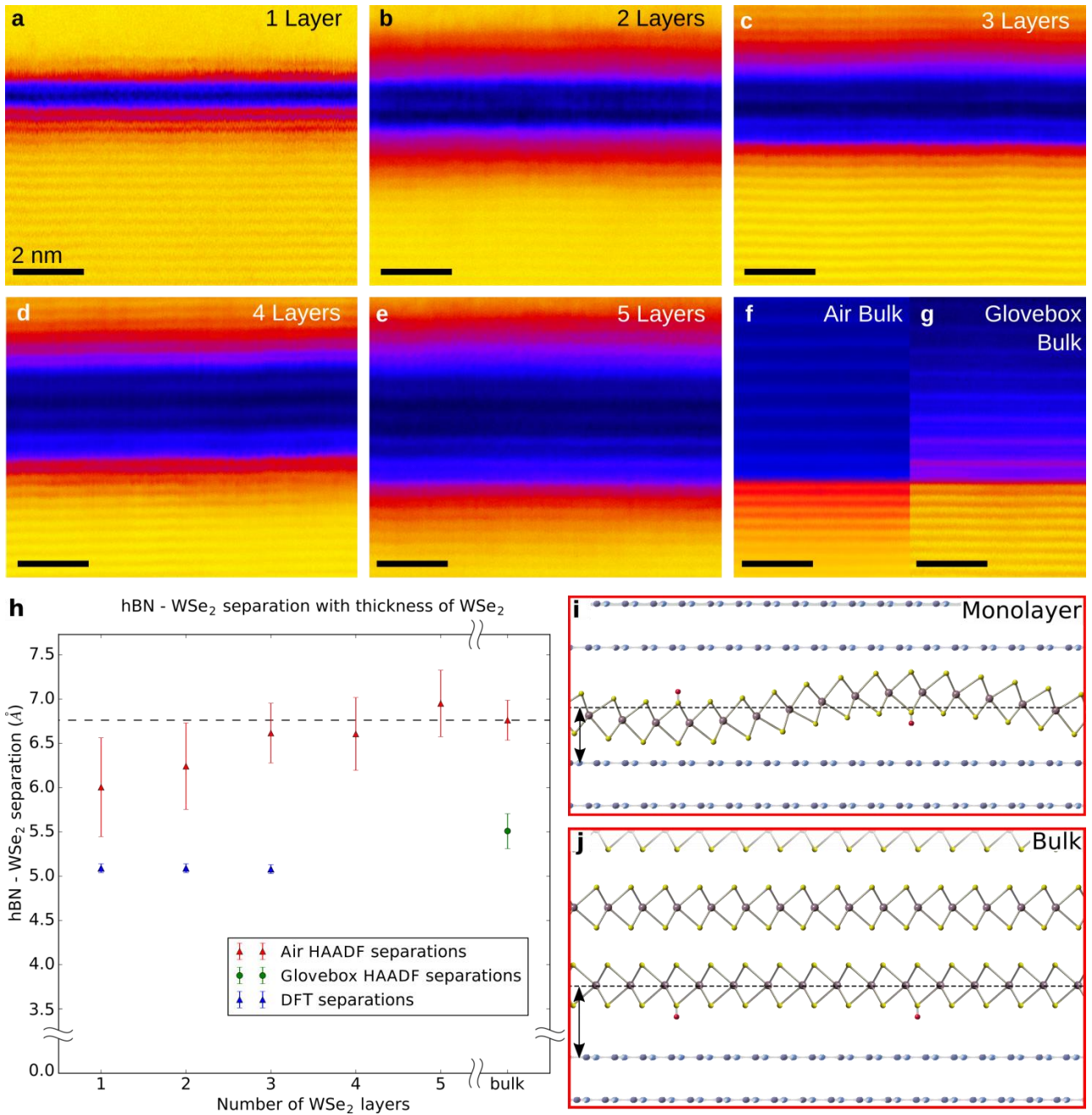

**Figure 4.** Cross sectional HAADF STEM analysis of different thicknesses of WSe$_2$ from monolayer up to 5 layer and bulk WSe$_2$, encapsulated in hBN. **a,** a monolayer of WSe$_2$ encapsulated by hBN. **b – e** are near identical systems to **a** but with 2, 3, 4 and 5 layer-thick 2H-stacked WSe$_2$ in the place of the monolayer. **f** shows a single interface between bulk WSe$_2$ (~50 layers) and hBN. **g** is a similar interface between bulk WSe$_2$ and hBN but fabricated in an argon gas environment using a glovebox. **f** and **g** are aligned by their hBN lattices (yellow/red) and the offset in the respective WSe$_2$ lattices (dark blue) arises from different interface distances. All scale bars 2 nm. **h** Quantitative analysis of the HAADF images allows us to determine the interlayer distance at the interface between hBN and WSe$_2$. The values as calculated by DFT are plotted blue and the experimentally determined values for air (glovebox) fabrication are plotted

as red triangles (green circles). Error bars represent standard deviations in the measurements and the dashed line denotes the bulk value for air fabrication. DFT predicts no change in the interlayer separation for different thicknesses of encapsulated WSe$_2$. The hBN-WSe$_2$ measurements in air are all ~1 – 2 Å larger than the DFT values for pristine interfaces, while fabrication in Ar reduces this distance to within 0.5 Å of the DFT values. **i** Schematic demonstrating the effect of bending modulus on the measured interlayer separation for a defective monolayer TMDC flake, which bends to accommodate a disperse number of oxygen defects in the structure. The flake can deform around the defect sites resulting in an overall reduction in the average distance between the transition metal plane and the nearest hBN plane. **j** a bulk TMDC flake is too rigid to bend around the defect sites, resulting in a larger measured interlayer separation. This accounts for the trend in experimental interlayer separation measurements in **h**. In both schematics, the mean position of the transition metal atoms is denoted by a dashed black line and the corresponding measured interlayer separation is annotated with a double headed arrow.

**Methods:**

Device fabrication

The bottom hBN is mechanically exfoliated on a Si/SiO$_2$ wafer, whilst the TMDC and top hBN flakes are exfoliated onto separate sheets of thin PMMA. The TMDC is then pressed onto the bottom hBN layer and peeled from the PMMA (dry transfer). Top hBN is transferred onto the TMDC in the same way. The PMMA membrane is cut and washed away with acetone, leaving an encapsulated heterostructure.[10,24,25]

Cross sectional specimen preparation

Cross sections of these heterostructures were prepared using a dual beam instrument (FEI Dual Beam Nova 600i) combining a focused ion beam (FIB) and a scanning electron microscope (SEM). Prior to FIB milling a heterostructure device is coated with ~10 nm of carbon and ~3 nm of Au/Pd via sputtering. A suitable area is found using the SEM capabilities of the instrument. Flake edges and bubbles are still visible even after coating, allowing accurate, site specific cross-sections to be made. The *in situ* lift-out procedure[10,27] used a gas-injection system and the electron and ion beams to locally deposit a 15x1x1 µm protective platinum 'strap' across the surface of the device, defining the lamella geometry in plan view. FIB milling using 30 kV gallium ions is used to dig trenches either side of the strap and cut the resulting lamella free from the substrate using decreasing current steps of 9.3 – 1 nA. A micromanipulator needle is used to remove the lamella from the trench and transfer it to an Omniprobe copper half grid where it is secured by further Pt deposition. Low energy ion polishing (5 kV and 2 kV at 80 pA) was used to remove side damage and thin the lamella to 30 – 70 nm thickness.

STEM imaging

High resolution STEM imaging was carried out using a probe side aberration-corrected FEI Titan G2 80-200 kV with an X-FEG electron source. Bright field and high angle annular dark field (HAADF) imaging were performed using a probe convergence angle of 21 mrad, a HAADF inner angle of 48 mrad and a probe current of ~ 75 pA. The lamellae were aligned with the basal planes parallel to the incident electron probe but away from a low index zone axis in any individual layer. Images were acquired with the scan direction perpendicular to the atomic layers, so as to largely eliminate artefacts associated with specimen drift across the interfaces. Correct identification of each atomic layer within bright field and HAADF images was achieved by elemental analysis using either energy dispersive x-ray spectroscopy (EDXS) or electron energy loss spectroscopy (EELS). Where necessary, post processing alignment procedures were applied to compensate for any specimen drift that may have occurred by using the assumption that planes in the bulk hBN far from the interface were atomically flat. HAADF intensity profiles were acquired perpendicular to the atomic fringes and principle component analysis was used to denoise each profile (further information in supplementary information section 2). A peak finding algorithm searching for local maxima was applied to identify the position of lattice planes in the intensity profiles.

DFT

Density functional theory (DFT) calculations were performed using the projected augmented-wave method as implemented in VASP 5.3.5. The exchange–correlation functional is described by the revised Perdew–Burke–Ernzerhof (PBE) exchange model with the empirical dispersion correction of optB88-vdW (Becke88 van der Waals) functional.[23] For more details see supplementary information section 1.

ASSOCIATED CONTENT

**Supporting Information**

The following files are available free of charge.

Supporting Information for Observing imperfection in atomic interfaces in for van der Waals heterostructures (PDF):

- 1. Density functional theory calculations

- 2. HAADF STEM image processing using principle component analysis

- 3. Spectrum imaging of $WSe_2$ heterostructures

AUTHOR INFORMATION

**Corresponding Author**

Dr Sarah Haigh, School of Materials, University of Manchester, Oxford Road, Manchester, M13 9PL, United Kingdom. Sarah.haigh@manchester.ac.uk

Dr Roman Gorbachev, National Graphene Institute, University of Manchester, Oxford Road, Manchester, M13 9PL, United Kingdom. roman@manchester.ac.uk

**Author Contributions**

APR, RG and SJH wrote the manuscript. AK, MJH, FW, YC and RG fabricated the devices. APR and EP performed the STEM characterization. ANR and MIK performed DFT calculations.

All authors contributed to the interpretation of the results and have given approval to the final version of the manuscript.

**Funding Sources**

This work was supported by the UK Engineering and Physical Sciences Research Council (EPSRC) (grant numbers EP/K016946/1 and EP/M010619/1) and the Royal Society UK. SJH and APR also acknowledge support from the US Defense Threat reduction agency (grant HDTRA1-12-1-0013), EPSRC NowNano EPSRC doctoral training center and ERC Starter grant EvoluTEM.

ACKNOWLEDGMENT

The authors are grateful to Prof Gordon Tatlock and Dr Simon Romani for access to the aberration corrected 2100F at the University of Liverpool.

ABBREVIATIONS

2D materials, two-dimensional materials; DFT, density functional theory; FIB, focused ion beam; HAADF, high angle annular dark field; hBN, hexagonal boron nitride; (S)TEM, (scanning) transmission electron microscopy; SEM, scanning electron microscope; TMDC, transition metal dichalcogenide.

# Supporting Information for:

# Observing imperfection in atomic interfaces for van der Waals heterostructures

*Aidan. P. Rooney[1], Aleksey Kozikov[2,3,4], Alexander N. Rudenko[5], Eric Prestat[1], Matthew J Hamer[2,3,4], Freddie Withers[6], Yang Cao[2,3,4], Kostya S. Novoselov[2,3,4], Mikhail I. Katsnelson[5], Roman Gorbachev[2,3,4]\* Sarah J. Haigh[1,4]\**

1. School of Materials, University of Manchester, Manchester M13 9PL, UK

2. Manchester Centre for Mesoscience and Nanotechnology, University of Manchester, Manchester M13 9PL, UK

3. School of Physics and Astronomy, University of Manchester, Oxford Road, Manchester, M13 9PL, UK

4. National Graphene Institute, University of Manchester, Manchester M13 9PL, UK

5. Institute for Molecules and Materials, Radboud University, 6525 AJ Nijmegen, Netherlands

6. College of Engineering, Mathematics and Physical Sciences, University of Exeter, Exeter, Devon, EX4 4SB, UK

1. **Density functional theory calculations**

The experimental HAADF STEM images were compared to density functional theory (DFT) calculations in order to help explain the observed structures.

DFT calculations were carried out using the projected augmented-wave method (PAW)[1] as implemented in the Vienna ab initio simulation package (VASP)[2,3]. Exchange and correlation effects were taken into account within the dispersion-corrected nonlocal vdW-DF functional[4] in the parametrization of Klimeš *et al.* (optB88-vdW)[5], which shows good performance for weakly-bonded layered solids including graphene and h-BN[6]. An energy cutoff of 550 eV for the plane-wave basis and the convergence threshold of $10^{-6}$ eV were used in the self-consistent solution of the Kohn-Sham equations, which have proven to be sufficient to obtain numerically converged forces to within $10^{-2}$ eV/Å. For the 2$^{nd}$ (3$^{rd}$) row transition metals (TMs) the 5$s$ and 4$d$ (6$s$ and 5$d$) electrons only were treated as valent. The inclusion of the 4$p$ (5$p$) electrons did not affect the interlayer separations by more than 0.01 Å. For the *p*-elements, only $s$ and $p$ electrons of the outer shell were treated as valent. The encapsulated systems were modelled in the bulk geometry with the unit cell containing one or two layers of TMDC and three layers of hBN in the AA' stacking. The atomic structure and lattice parameters were fully relaxed. To minimize the lattice mismatch between the hBN and TMDC down to approximately 1% (see Table S1) we use hexagonal unit cells with the following lattice parameters for $MoS_2$, $WS_2$, $NbSe_2$, $MoSe_2$, and $WSe_2$, respectively: 5$a$, 5$a$, 7$a$, 8$a$, 8$a$, where $a \sim 2.50$ Å is the lattice parameter of h-BN. The Brillouin zone was sampled by a uniform distribution of 64 or 16 **k**-points, depending on the unit cell dimensions.

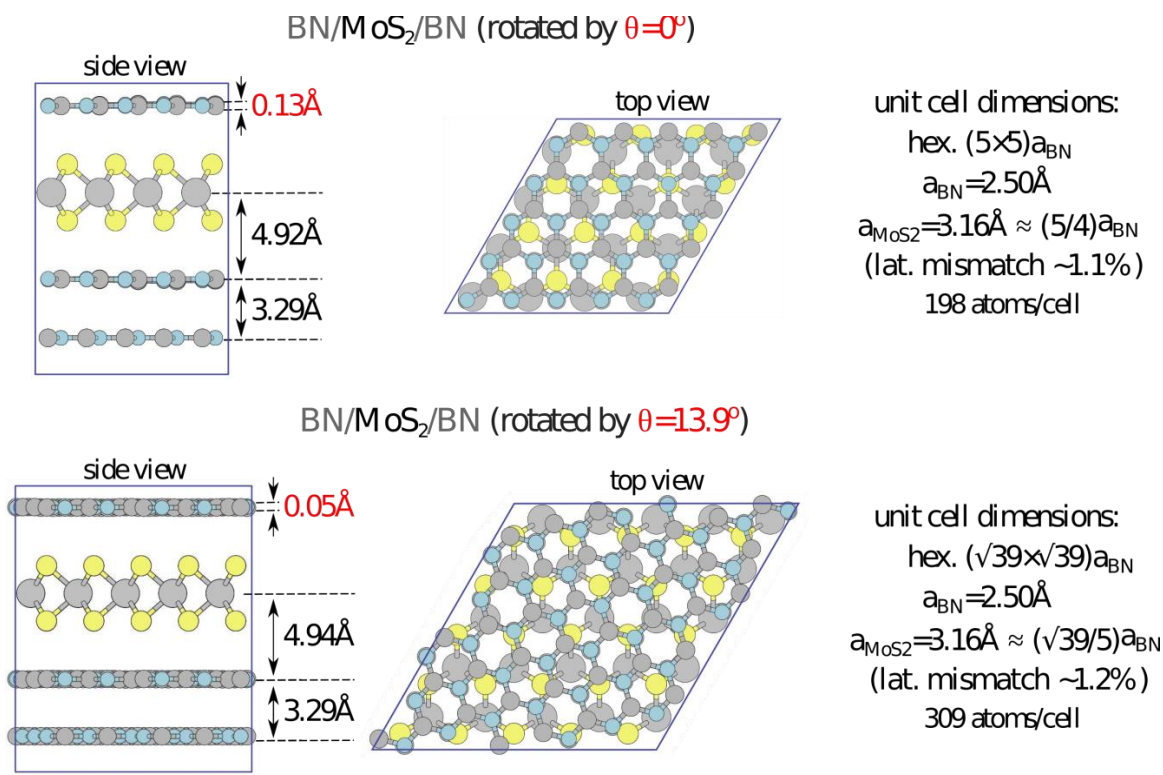

**Figure S1.** Effect of twist angle on interlayer separation in hBN/MoS$_2$/hBN. **Top:** Side and top views of hBN encapsulated MoS$_2$ with a twist angle of 0.0°. **Bottom:** Side and top views of hBN encapsulated MoS$_2$ with a twist angle of 13.9°. Both fully relaxed structures show negligible differences in interlayer separation, hBN spacing and variation.

In order to assess the accuracy of the DFT calculations and reduce the number of computationally expensive simulations to be run, the effect of lattice twist angle on distances in the heterostructure system was investigated. A monolayer of MoS$_2$ encapsulated in hBN was orientated with azimuthal twist angles of 0° and 13.9° and both systems relaxed (see Fig. S1). The bulk spacing of hBN, the interlayer separation between MoS$_2$ and hBN, and the variation in hBN vertical positions differed by ≤0.08 Å, which is significantly less than the accuracy of our experimental data (± 0.5 Å). Therefore, we assume the effect of twist angle on the measured distances of our system is negligible.

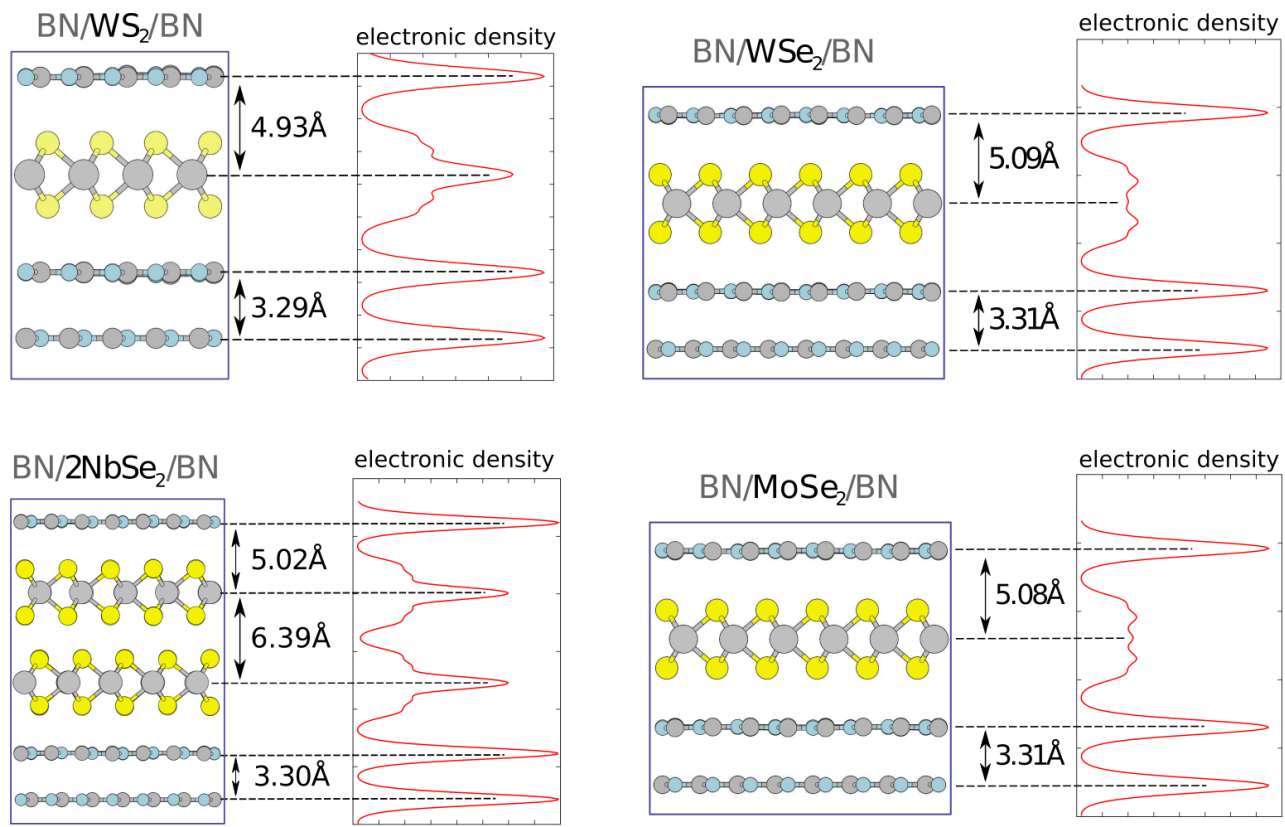

**Figure S2.** Interlayer separation in hBN/TMDC/hBN systems as predicted by DFT. $WS_2$, $MoSe_2$ and $WSe_2$ monolayers and $NbSe_2$ bilayer flakes trapped in hBN. Important distances are annotated on each system. All interlayer distances are predicted to be between 4.92 – 5.09 Å.

The calculation parameters used for the $hBN/MoS_2/hBN$ system were applied to monolayer and bilayer TMDC systems with twist angles of 0°. Between all systems the bulk hBN spacing varies by 0.02 Å, whilst the interlayer distances lie between 4.92 – 5.09 Å. These values do not explain the discrepancies observed in our experimental results, leading us to question whether differences in chemical reactivity can explain why $MoSe_2$ and $WSe_2$ deviate from the expected values. Table S2 shows DFT calculated values for the work function and electron affinity for each TMDC. The energy terms used to calculate these properties are $E_{VBM}$ and $E_{cbm}$, corresponding to the valence band and conduction band minima, and $E_{vac}$ corresponding to vacuum. Low values of work function predict a higher reactivity with acceptor species (behaves as a nucleophile), whilst high values of electron affinity predict a higher reactivity with donor species (behaves as an electrophile). From this we would expect MoSe$_2$ and WSe$_2$ to be the most nucleophilic TMDCs, however the values are not extreme enough to warrant prediction of substantial reaction with contaminant species trapped between the layers.

**Table S1**: Lattice mismatches ($\Delta a=(a_{exp}-a_{TMDC})/a_{TMDC}$) between the experimental TMDC lattice constants ($a_{exp}$) and lattice constants used for DFT simulations of BN/TMDC/BN heterostructures ($a_{TMDC}$). Experimental TMDC lattice constants are taken from Ref. 7. $N_{at}$ is the number of atoms in the simulated supercell.

| TMDC | $a_{TMDC}/a_{BN}$ | $a_{exp}$, Å | $\Delta a$, % | $N_{at}$ |
|---|---|---|---|---|
| MoS$_2$ | 5/4 | 3.16 | +1.1 | 198 |
| MoS$_2$* | √39/5 | 3.16 | +1.2 | 309 |
| WS$_2$ | 7/5 | 3.15 | +0.8 | 198 |
| NbSe$_2$ | 7/5 | 3.44 | −1.7 | 444 |
| MoSe$_2$ | 8/6 | 3.29 | −1.3 | 492 |
| WSe$_2$ | 8/6 | 3.28 | −1.6 | 492 |

* Rotated by 13.9°

**Table S2**
**Quantities related to the reactivity of TMDC: work function and electron affinity**

|  | MoS$_2$ | WS$_2$ | MoSe$_2$ | WSe$_2$ | NbSe$_2$ |
|---|---|---|---|---|---|
| Work function* = $E_{VBM} - E_{vac}$ | 5.93 eV | 5.70 eV | 5.33 eV | 5.10 eV | 5.55 eV |
| Electron affinity** = $E_{CBM} - E_{vac}$ | 4.27 eV | 3.94 eV | 3.88 eV | 3.56 eV | 5.55 eV |

\* lower values mean higher reactivity wrt acceptor species (like O$_2$)
\*\* higher values mean higher reactivity wrt donor species (like metals)

**Simulating defects at the interface**

To better understand the extent to which an impurity may disrupt the interfaces for a van der Waals heterostructure, an OH molecule was added to the van der Waals gap between the upper hBN and the encapsulated monolayer WSe$_2$ as shown in Figure S3 below. The calculation was run identically to those presented in Figure S2 with the exception of the added impurity.

We find the separation of the upper hBN from the WSe$_2$ monolayer increases by 0.68 Å to 5.77 Å (5.09 Å in the pristine case) when an OH impurity is placed at the interface. In addition the separation of the lower hBN from the WSe$_2$ monolayer decreases slightly (to 4.97 Å). This theoretical result compares well to our measurement in the main text of 6.01 Å ± 0.56 Å. Nevertheless, limitations on the size of the simulation that can be performed and the use of periodic boundary conditions prevent us from modelling corrugations to the monolayers or accurately reproducing the low defect densities we observe experimentally. As a result this comparison is qualitative at best.

It should be noted that previous DFT calculations have similarly predicted an increase in interlayer separation between graphene and silicon when the interface is perturbed by hydrogen (3.3 Å increase) or water (3.5 Å) impurities[8,9], and for chemisorbed oxygen on a various TMDCs the oxygen-metal distances were found to be in the range of 2 Å[10].

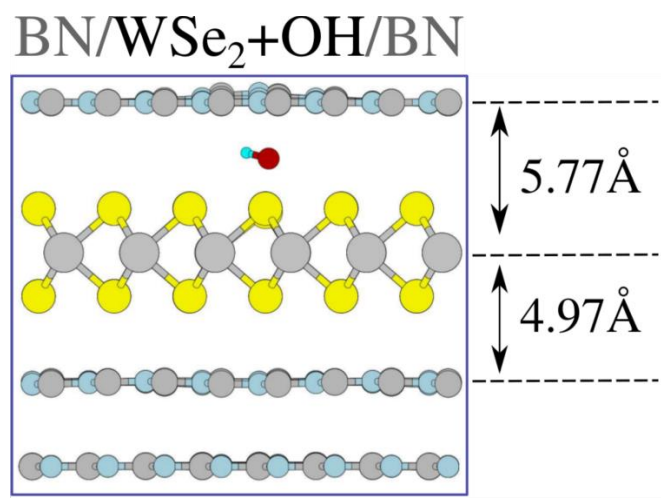

**Figure S3.** Interlayer separation in hBN/ WSe$_2$ + OH/ hBN systems as predicted by DFT. Important distances are annotated. The interlayer distance is increased from the pristine case by 0.68 Å to 5.77 Å, whilst the unperturbed side is decreased to 4.97 Å.

## 2. HAADF STEM image processing using principle component analysis

To accurately determine the lateral positions of the different 2D crystals layered within in a van der Waals heterostructure, the high angle annular dark field (HAADF) scanning transmission electron microscopy (STEM) image needs to be carefully processed. Fig. S4A shows an atomic schematic of the structure being imaged in Fig. S4C (this image was first aligned using hBN fringes to remove scan noise and specimen drift artefacts). To quantitatively measure the interlayer spacing between two dissimilar crystals (hBN – TMDC) an intensity profile is acquired perpendicular to the lattice fringes, each profile having a width of one pixel (0.4 Å). A typical intensity profile taken from the raw image is shown in Fig. S4D, revealing that, although the peaks are visible, the exact positions of their maxima are not readily determined.

Principle component analysis (PCA) is a statistical algorithm which can be used to separate the components of a dataset by their degree of variability.[11] By assigning the axis parallel to the basal planes as the navigation axis, and the axis perpendicular to the basal planes (i.e. the direction of the intensity profile) as the signal axis, PCA can distinguish the components which comprise the raw intensity profile by their variance. A scree plot showing the explained variance ratio of each of principle component of the raw image is presented in Fig. S4B. As might be expected the components with the largest variation correspond to the main features in the image: in this case the peaks and troughs of the atomic planes in the heterostructure, whilst the components with the smallest variation are composed entirely of noise. Components between these two extremes hold some useful signal and some noise signal. As such, a judgement is made as to how many components to include in the reconstructed image; in this scenario we have chosen 21 components. Fig. S4E shows the reconstructed image, which compares favorably to the original. The processed intensity profile in Fig. S4F now shows smooth peaks with readily identifiable maxima. This allows measurement of the distance between nearest neighbor fringes from these maxima with high precision. A peak finding algorithm[12], searching for local maxima can then be successfully applied to accurately identify the position of lattice planes.

Applying PCA to uniform structures shown in this work is very useful for denoising images and intensity profiles, however we note that it may prove to be a key method for analyzing images showing tortuous interfaces or heterogeneous features.

It should also be noted the different crystal lattices are azimuthally rotated to one another (sometimes referred to as 'twist angle'). The lamella can be rotated such that each crystal comes 'on zone' and atomic resolution is observed from orientated lattices, as shown in Figure S5. This is accompanied by enhanced channeling contrast which can interfere with measurement accuracy. Therefore, to get equivalent signals from all lattices the lamellae were tilted to an angle where channeling contrast was minimized and the basal planes were aligned to the incident electron probe. This improved the success of the automated peak-finding and the accuracy of the interlayer separation measurement. The measured interlayer separations for different thicknesses and fabrication methods for $WSe_2$/hBN are presented in Table S3.

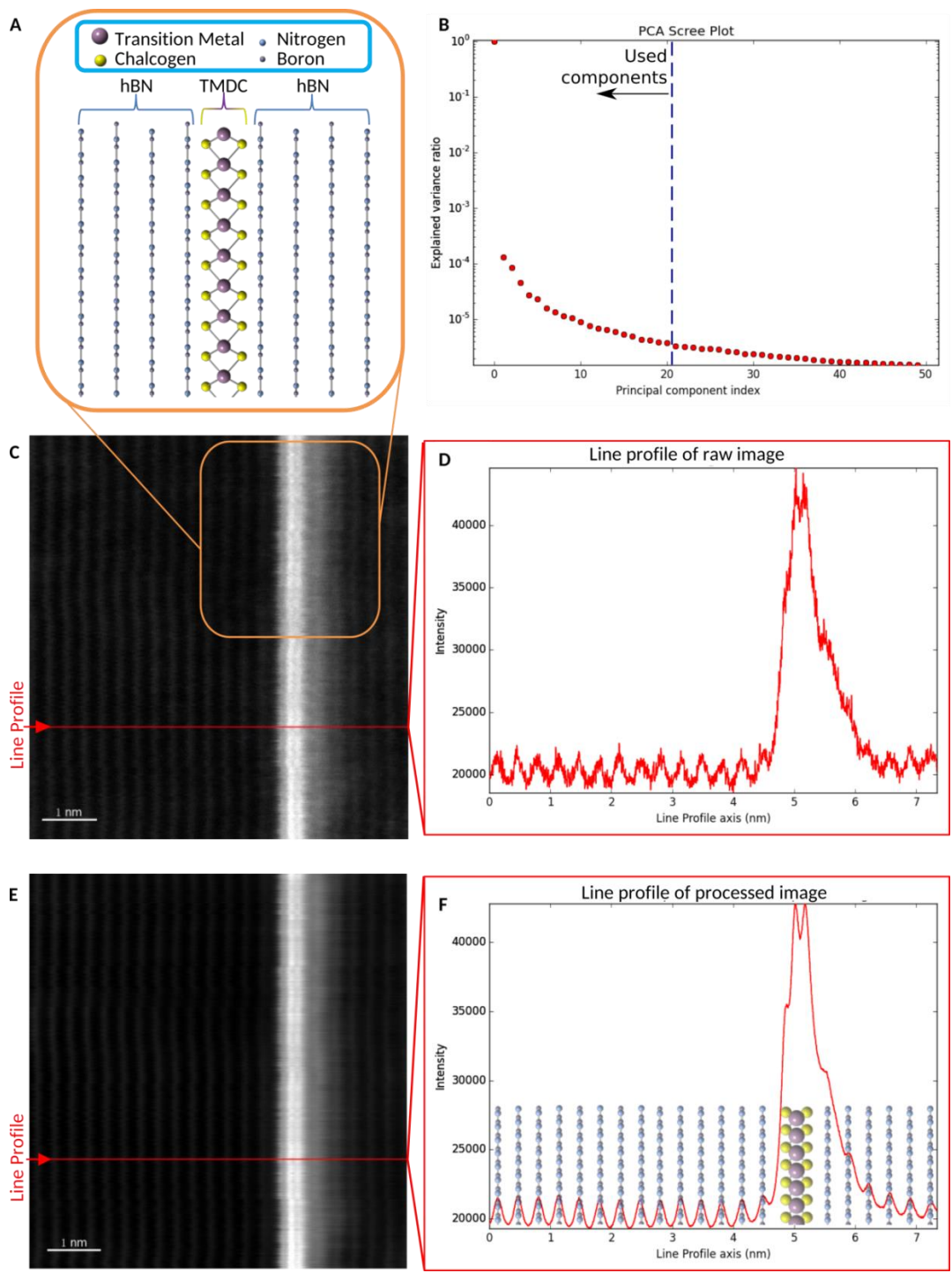

**Figure S4.** Denoising similar intensity profiles in HAADF STEM images. **A**, Schematic showing the positions of atoms in the highlighted region shown in **C**. A monolayer TMDC is encapsulated by bulk hBN. An example intensity profile **D** shows how noise prevents us from successfully identifying peaks which correspond to atomic positions. **B** The explained variance ratio of each component, with the threshold annotated. **E** The denoised image shows all the same features of its parent in **C**, however the intensity profile in **F** is now fully denoised and we can assign atomic components to each peak.

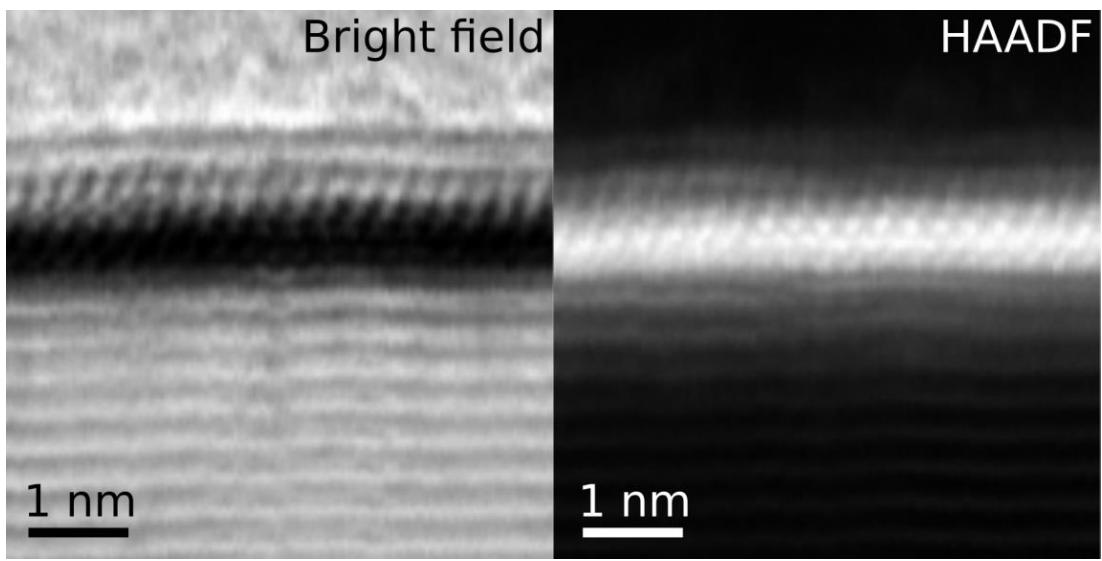

**Figure S5. Left** Bright field and **Right** HAADF STEM images of the same region. In this case the WSe$_2$ monolayer is aligned to the incident probe and the three rows of atomic columns (Se – W – Se) are resolved. Due to random azimuthal rotation of the encapsulating hBN they are not on zone and are seen instead as continual planes of atoms. For all other images considered in this work, the lamellae were tilted to this off-zone condition to give more accurate measurements and avoid dominant channeling contrast.

**Table S3**: Table of measured interlayer separations for 1-5L and bulk WSe$_2$.

| Thickness of WSe$_2$ | Mean hBN – W distance (Å) | Standard Deviation (Å) |
|---|---|---|
| Monolayer | 6.01 | 0.56 |
| Bilayer | 6.24 | 0.49 |
| Trilayer | 6.62 | 0.34 |
| Four layer | 6.61 | 0.41 |
| Five layer | 6.95 | 0.38 |
| Bulk – Air | 6.76 | 0.23 |
| Bulk - Glovebox | 5.51 | 0.20 |

**Roughness analysis of buried interfaces**

The perturbation of the van der Waals interface can be further quantified by measuring the roughness of the TMDC lattice fringe(s) relative to the nearest neighbor hBN fringe. This is captured as variation in the measurement of interlayer separation together with measurement error. This can be seen in the example datasets from 1-5 layer and bulk $WSe_2$ are presented in Fig. S6.

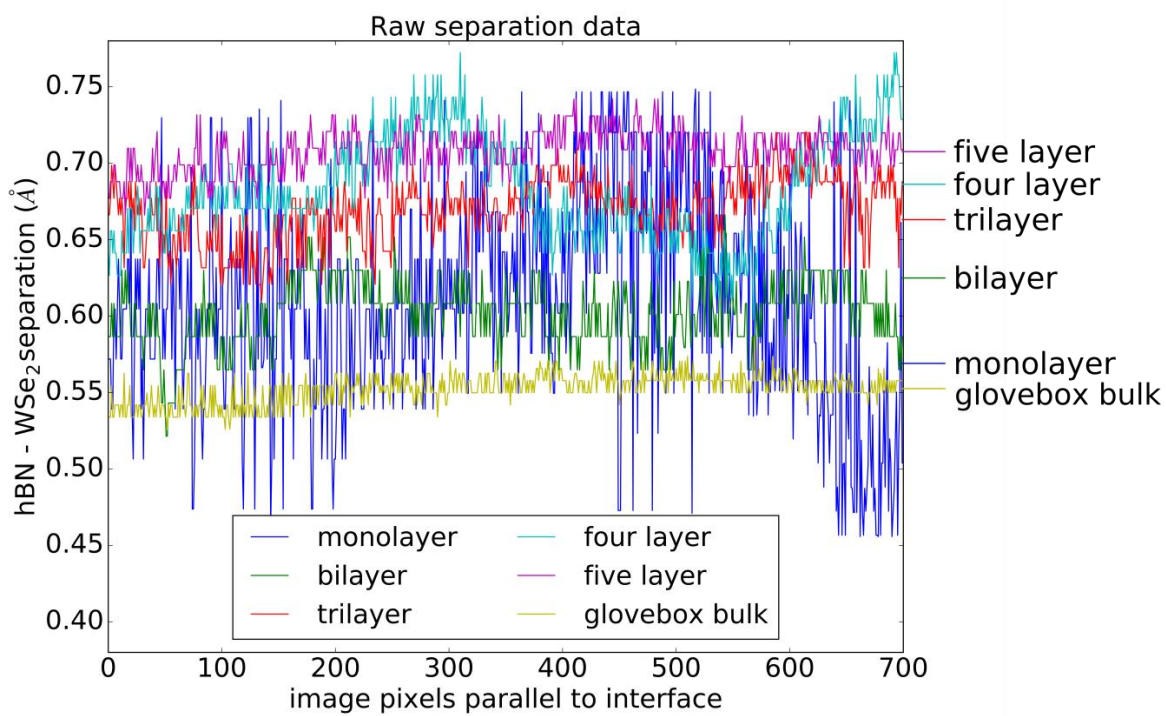

**Figure S6.** Unprocessed separation measurement data for different thicknesses of WSe2.

The roughness average for 1-5 layer and glovebox fabricated bulk $WSe_2$ is plotted in Fig. S7. The roughness was found using the roughness average ($R_a$):

$$R_a = \frac{1}{L}\int_0^L |Z(x)|\, dx$$

Where L is the distance over which the measurement (10 – 15 nm) was carried out and Z is the height at position x. As expected, thin flakes of $WSe_2$ are rougher than thicker flakes as the materials become stiffer with thickness.

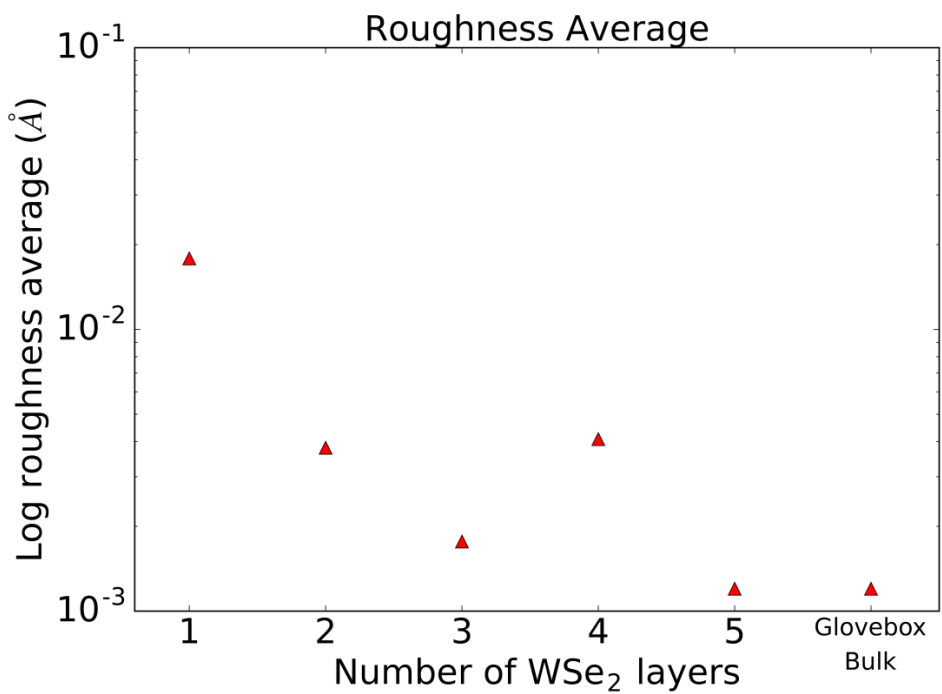

**Figure S7.** A log-plot of roughness average calculated from the separation measurements of 1-5 layer and glovebox bulk $WSe_2$. The thinner flakes show enhanced roughness as they deform around defects at the interface with hBN. Thicker flakes are stiffer and so less rough.

3. **Spectrum imaging of WSe$_2$ heterostructures**

Energy dispersive X-ray spectroscopy (EDXS) and electron energy loss spectroscopy (EELS) analysis was used to characterize the chemistry of the buried interfaces in WSe$_2$. The mailtin motivation for this was to determine if the interfaces between WSe$_2$ and hBN had been chemically modified, resulting in an increased interlayer separation at this interface. Fig S8 shows the distribution of elements through the depth of the structure. Tungsten, selenium and nitrogen distributions confirm the successful encapsulation of the WSe$_2$. The background subtracted EELS carbon and nitrogen K-edge intensities plotted across the same region can be used to determine if contamination is present at these two interfaces either side of the WSe$_2$ monolayer. The carbon distribution clearly identifies the thick amorphous carbon coating on the surface of the upper hBN. The interfaces, in contrast, show no discernible presence of carbon. Similarly, oxygen shows no signal at the interfaces, only a small enhancement at the surface of the upper hBN. From this we conclude that any chemical modification of the WSe$_2$ is at a level below the detection limit of this technique (~1 at%).

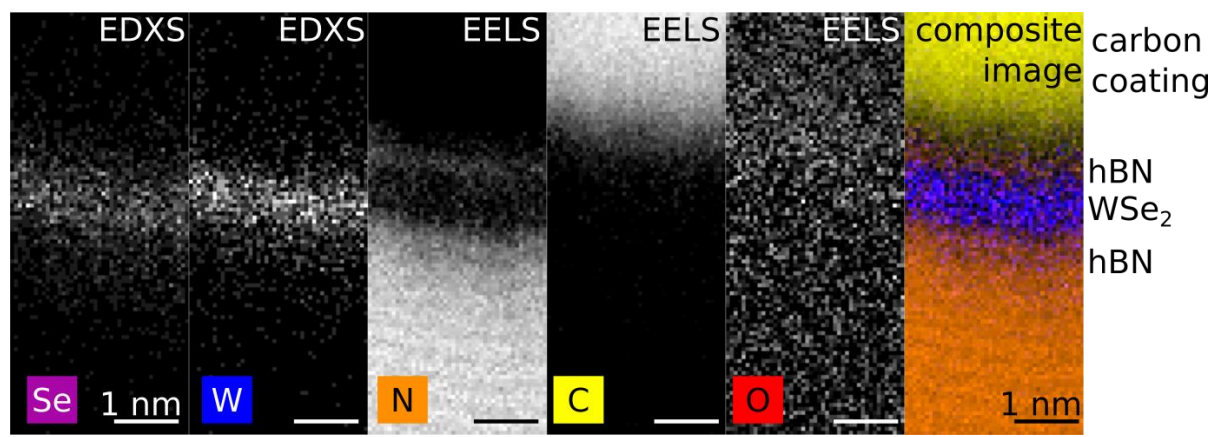

**Figure S8.** EDXS and EELS spectrum imaging, mapping the distribution of selenium, tungsten, nitrogen and carbon in a $WSe_2$ monolayer encapsulated in hBN**.** EDXS maps are formed from expressing the peak intensities as greyscale values for every pixel in the spectrum image. Maps for the selenium K-α and tungsten L-α characteristic X-ray peaks are shown leftmost. EELS maps are formed from plotting the intensities of background subtracted characteristic edges. The nitrogen, carbon and oxygen K-edges are plotted center. Rightmost is the composite image of all four maps, with selenium plotted magenta, tungsten purple, nitrogen orange, carbon yellow. No carbon or oxygen is seen the interfaces between hBN and $WSe_2$, despite the presence of an amorphous carbon coating on the surface of the structure to aid electron microscope sample preparation. The apparent bending of the basal planes is due to specimen drift during the long acquisition time used to maximize sensitivity. All scale bars 1 nm.